\documentclass[preprint,aps,prb]{revtex4}
\usepackage[dvips]{graphicx}
\usepackage{color,array,dcolumn}
\usepackage{amsmath}
\usepackage{amssymb}

\begin{document}

\title{Van der Waals Interactions in DFT using Wannier Functions
without empirical parameters}
\author{Pier Luigi Silvestrelli}
\affiliation{Dipartimento di Fisica e Astronomia ''G. Galilei'', 
Universit\`a di Padova, via Marzolo 8, I--35131, Padova, Italy, 
and CNR-IOM Democritos, via Bonomea 265, I--34136, Trieste, Italy.}
\author{Alberto Ambrosetti}
\affiliation{Dipartimento di Fisica e Astronomia ''G. Galilei'', 
Universit\`a di Padova, via Marzolo 8, I--35131, Padova, Italy, 
and CNR-IOM Democritos, via Bonomea 265, I--34136, Trieste, Italy.}

\begin{abstract}
\date{\today}
A new implementation is proposed for including van der Waals (vdW) interactions
in Density Functional Theory (DFT) using the Maximally-Localized Wannier 
functions (MLWFs), which is free from empirical parameters. 
With respect to the previous DFT/vdW-WF2 method, in the
present DFT/vdW-WF2-x approach, the empirical, short-range, damping 
function is replaced by an estimate of the Pauli exchange repulsion,
also obtained by the MLWFs properties.
Applications to systems contained in 
the popular S22 molecular database and to the case of an Ar atom interacting 
with graphite, and comparison with reference data,
indicate that the new method, besides being more physically founded, 
also leads to a systematic improvement in the description 
of vdW-bonded systems.
\end{abstract}

\maketitle
                                                                                
\section{Introduction}
Density Functional Theory (DFT) is a well-established
computational approach to study
the structural and electronic properties of
condensed matter systems from first principles.
Although current, approximated density functionals
allow a quantitative description
at much lower computational cost than other
first principles methods, they fail\cite{Kohn} to properly describe
dispersion interactions. Dispersion forces originate from
correlated charge oscillations in
separate fragments of matter and the most important
component is represented by the $R^{-6}$ van der Waals (vdW)
interaction,\cite{london} due to correlated instantaneous dipole
fluctuations.
These interactions play a fundamental role in determining the
structure, stability, and function of a wide variety of systems,
including molecules, clusters, proteins, nanostructered materials,
molecular solids and liquids, and in adsorption processes of fragments weakly
interacting with a substrate (''physisorbed'').

In the last few years a variety of practical methods have been proposed
to make DFT calculations able to accurately describe vdW effects (for a
review, see, for instance, refs. 
\onlinecite{Riley,MRS,Klimes,Woods,Grimme2016,Hermann}).
In this respect, a family of such methods, all based
on the Maximally Localized Wannier Functions (MLWFs),\cite{wannier}
has been developed, namely the original
DFT/vdW-WF scheme,\cite{silvprl,silvmetodo,Mostofi}
DFT/vdW-WF2\cite{C3} (based on the London expression and taking into account
the intrafragment overlap of the MLWFs), 
DFT/vdW-WF2s\cite{PRB2013} (including metal-screening corrections), and
DFT/vdW-QHO-WF\cite{myQHO} 
(adopting the coupled Quantum Harmonic Oscillator model),
successfully applied to a variety of systems:
\cite{silvprl,silvmetodo,silvsurf,CPL,silvinter,myQHO,ambrosetti,C3,Ar-Pb,Costanzo,Ambrosetti2013,PRB2013,Mostofi}
small molecules, water clusters,
graphite and graphene, water layers interacting with graphite,
interfacial water on semiconducting substrates,
hydrogenated carbon nanotubes,
molecular solids, the interaction of rare gases and small molecules
with metal surfaces,...
 
In all these methods a certain degree of empiricism is present
since the energetic vdW-correction term is multiplied by a short-range
{\it damping function}, which is
introduced not only to avoid the unphysical divergence of the
vdW correction at small fragment separations, but also
to eliminate double countings of correlation effects 
(in fact standard DFT approaches are able to describe short-range 
correlations). This damping function contains one or more empirical
parameters which are typically set by a trial and error approach
or/and are fitted using some reference database.

In the present paper we overcome the above limitation by presenting 
a new method, hereafter referred to as DFT/vdW-WF2-x, where
the empirical, short-range, damping function is replaced by an estimate 
of the Pauli {\it exchange} repulsion, also obtained by the MLWFs properties.
The new approach is successfully 
applied to the popular S22 benchmark set\cite{Jurecka} of weakly interacting
molecules and also to the case of an Ar atom interacting with 
graphite.
The results are compared with 
reference data and indicate that the new method leads to a systematic 
improvement in the description of vdW-bonded systems.

\section{Method}
Here we describe and apply a new implementation of the DFT/vdW-WF2
method, introduced in ref. \onlinecite{C3} and summarized below. 
Basically, the electronic charge partitioning
is achieved using the Maximally-Localized Wannier Functions (MLWFs), which
are obtained from a unitary transformation in the space of the
occupied Bloch states, by minimizing the total spread functional:\cite{wannier}
 
\begin{equation}
\Omega=\sum_n S^2_n=\sum_n \left( <w_n|r^2|w_n>-<w_n|\mathbf{r}|w_n>^2\right).
\end{equation}

The localization properties of the MLWFs are of particular interest for 
the implementation of an efficient
vdW correction scheme: in fact, the MLWFs represent a suitable basis set
to evaluate orbital-orbital vdW interaction terms. 
In particular, if two interacting atoms, $A$ and $B$, are
approximated\cite{london} by coupled harmonic oscillators, the vdW
energy correction can be taken to be the change of the zero-point energy
of the coupled oscillations as the atoms approach; if only a single
excitation frequency is associated to each atom, $\omega_A$, $\omega_B$,
then

\begin{equation}
E^{London}_{vdW}=-\frac{3e^4}{2m^2}\frac{Z_A Z_B}{\omega_A \omega_B(\omega_A+\omega_B)}\frac{1}{R_{AB}^6}
\label{lond}
\end{equation}

where $Z_{A,B}$ is the total charge of A and B, and $R_{AB}$ is 
the distance between the two atoms ($e$ and $m$ are the electronic charge
and mass).
Now, adopting a simple classical theory of the atomic polarizability, 
the polarizability 
of an electronic shell of charge $eZ_i$ and mass $mZ_i$, tied to a heavy 
undeformable ion can be written as

\begin{equation}
\alpha_i\simeq \frac{Z_i e^2}{m\omega_i^2}\,.
\label{alfa}
\end{equation}

Then, given the direct relation between polarizability and atomic
volume,\cite{polvol} we assume that $\alpha_i\sim \gamma S_i^3$,
where $\gamma$ is a proportionality constant, so that the atomic volume is
expressed in terms of the MLWF spread, $S_i$.
Rewriting eq. \eqref{lond} in terms of the quantities defined above, 
one obtains an explicit, simple expression for
the $C_6$ vdW coefficient: 

\begin{equation}
C_{6}^{AB}=\frac{3}{2}\frac{\sqrt{Z_A Z_B}S_A^3 S_B^3 \gamma^{3/2}}
{(\sqrt{Z_B}S_A^{3/2}+\sqrt{Z_A}S_B^{3/2})}\,.
\label{c6}
\end{equation}

The constant $\gamma$ can then be set up by imposing that the exact value for 
the H atom polarizability
($\alpha_H=$4.5 a.u.) is obtained. This appears to be a physically sound 
choice since, in the H case, one
knows the exact analytical spread, $S_i=S_H=\sqrt{3}$ a.u.

In order to achieve a better accuracy, one must properly deal with
{\it intrafragment} MLWF overlap (we refer here to charge overlap, not to be
confused with wave functions overlap): 
in fact, the method is strictly
valid for nonoverlapping fragments only; now, while the overlap between the
MLWFs relative to separated fragments is usually negligible for all the
fragment separation distances of interest, the same is not true for the 
MLWFs belonging to the same fragment, which are often characterized by
a significant overlap. This overlap affects the effective orbital volume,
the polarizability, and the excitation 
frequency (see eq. \eqref{alfa}), thus leading to a quantitative effect on the
value of the $C_6$ coefficient. 
We take into account the effective change in volume due to intrafragment 
MLWF overlap by introducing a suitable reduction factor $\xi$
obtained by interpolating between the limiting cases of fully
overlapping and non-overlapping MLWFs.
In particular,
since in the DFT/vdW-WF2 method the $i$-th MLWF is approximated 
with a homogeneous charged sphere of radius $S_i$, then the 
overlap among neighboring MLWFs can be evaluated as the geometrical
overlap among neighboring spheres. 
To derive the correct volume reweighting factor for dealing with overlap
effects, we first consider the limiting case of two pairs (one for
each fragment) of completely overlapping MLWFs, which would be, for instance,
applicable to two interacting He atoms if each MLWF just describes the
density distribution of a single electron; then we can evaluate
a single $C_6$ coefficient using eq. \eqref{c6} with $Z_{A,B}=2$,
so that:
 
\begin{equation}
C_{6}^{AB}=\frac{3}{2}\frac{\sqrt{2}S_A^3 S_B^3 \gamma^{3/2}}
{(S_A^{3/2}+S_B^{3/2})}.
\label{c62}
\end{equation}

Alternatively, the same expression can be obtained by considering the
sum of 4 identical pairwise contributions (with $Z=1$), by 
introducing a modification of the effective volume in such a way to
take the overlap into account and make the global interfragment
$C_6$ coefficient equivalent to that in eq. \eqref{c62}.
This is clearly accomplished by replacing $S_{i}^3$ in eq. \eqref{c6} with 
$\xi S_{i}^3$, where $\xi = 1/2$.
This procedure can be easily generalized to multiple 
overlaps, by weighting the overlapping
volume with the factor $n^{-1}$, where $n$ is the number of overlapping MLWFs.
Finally, by extending the approach to partial overlaps,  
we define the {\it free} volume of a set of MLWFs belonging to a given fragment
(in practice three-dimensional integrals are evaluated by numerical sums
introducing a suitable mesh in real space) as:

\begin{equation}
V_{free}=\int d\mathbf{r}\, w_{free}(\mathbf{r})
\simeq \Delta r \sum_l w_{free}(\mathbf{r}_l)
\end{equation}
where $w_{free}(\mathbf{r}_l)$ is equal to 1 if 
$|\mathbf{r}_l-\mathbf{r}_i|<S_i$ for at least one of the
fragment MLWFs, and is 0 otherwise.

The corresponding {\it effective} volume is instead given by
\begin{equation}
V_{eff}=\int d\mathbf{r}\, w_{eff}(\mathbf{r})
\simeq \Delta r \sum_l w_{eff}(\mathbf{r}_l)\,,
\end{equation}
where the new weighting function is defined as
$w_{eff}(\mathbf{r}_l)=w_{free}(\mathbf{r}_l)\cdot n_w(\mathbf{r}_l)^{-1}$,
with $n_w(\mathbf{r}_l)$ that is equal to the number of MLWFs
contemporarily satisfying the relation $|\mathbf{r}_l-\mathbf{r}_i|<S_i$.
Therefore, the non overlapping portions of the spheres (in
practice the corresponding mesh points) will be
associated to a weight factor 1, those belonging to two spheres to
a $1/2$ factor, and, in general, those belonging to $n$ spheres to
a $1/n$ factor. 
The average ratio between the effective volume and the free volume 
($V_{eff}/V_{free}$)
is then assigned to the factor $\xi$, appearing in eq. \eqref{c6eff}.
Although in principle the correction factor $\xi$
must be evaluated for each MLWF and the calculations must be repeated 
at different fragment-fragment separations, our tests show that,
in practice, if the fragments are rather homogeneous all the $\xi$ factors
are very similar, and if the spreads of the MLWFs do not change 
significantly in the range of the interfragment distances of
interest, the $\xi$'s remain essentially constant; clearly, 
exploiting this behavior leads to a significant reduction 
in the computational cost of accounting for the intrafragment overlap.
We therefore arrive at the following expression for the $C_6$ coefficient:

\begin{equation}
C_{6}^{AB}=\frac{3}{2}\frac{\sqrt{Z_A Z_B}\xi_A S_A^3 \xi_B S_B^3 \gamma^{3/2}}
{(\sqrt{Z_B\xi_A} S_A^{3/2}+\sqrt{Z_A\xi_B} S_B^{3/2})}\,,
\label{c6eff}
\end{equation}

where $\xi_{A,B}$ represents the ratio between the effective and the 
free volume associated to the $A$-th and $B$-th MLWF.
The need for a proper treatment of overlap effects has been also 
pointed out by Andrinopoulos {\it et al.},\cite{Mostofi} who however 
applied a correction only to very closely centred WFCs.

Finally, in the original DFT/vdW-WF2 method, the vdW interaction energy was 
computed as:

\begin{equation}
E_{vdW}=-\sum_{i<j}f(R_{ij})\frac{C_6^{ij}}{R^6_{ij}} =
-\sum_{i<j}\frac{C_6^{ij}}{R^6_{ij}} +
\sum_{i<j}\left(1-f(R_{ij})\right)\frac{C_6^{ij}}{R^6_{ij}}\,\,,
\label{esumold}
\end{equation}
where $f(R_{ij})$ is a short-range damping function, which is
introduced not only to avoid the unphysical divergence of the
vdW correction at small fragment separations, but also
to eliminate double countings of correlation effects 
(in fact standard DFT approaches are able to describe short-range 
correlations); it is defined as:   
\begin{equation}
f(R_{ij})=\frac{1}{1+e^{-a(R_{ij}/R_s-1)}}\,.
\end{equation}

The parameter $R_s$ represents
the sum of the vdW radii $R_s=R_i^{vdW}+R_j^{vdW}$, with
(by adopting the same criterion chosen above for 
the $\gamma$ parameter)
\begin{equation}
R_i^{vdW}=R_H^{vdW}\frac{S_i}{\sqrt{3}}\,\,,
\end{equation}
where $R_H^{vdW}$ is the literature\cite{Bondi} (1.20 \AA) vdW radius of 
the H atom, and, following Grimme {\it et al.},\cite{Grimme} 
$a \simeq 20$. Although $a$ is the only ad-hoc parameter of the method,
while all the others are only determined by the basic information given by 
the MLWFs (that is from first principles calculations) and in many 
applications the results are only mildly dependent on the particular value 
of $a$, nonetheless, this parameter, together with the choice of a 
specific form of the above damping function, clearly imply a certain 
degree of empiricism.

In order to overcome this limitation, we propose to improve the approach
by replacing the somehow artificial, short-range damping function by a 
term that directly measures the quantum mechanical Pauli {\it exchange} 
repulsion between electronic orbitals and
can be entirely expressed in terms of the MLWFs properties, without the
need of introducing empirical parameters.
Following ref. \onlinecite{Fedorov}, using the dipole approximation for
the Coulomb interaction, the exchange integral, for two closed electronic
shells with total zero spin, is simply given by:

\begin{equation}
J_{ex} = \frac{q^2 O}{2R}\,,
\end{equation}

where $q$ indicates the electronic charge of each electronic shell and
$O$ is the overlap integral between the electronic shells, separated by
$R$.
In our specific case, assuming that an electronic orbital is described
by the wave function relative to a {\it quantum harmonic oscillator}:

\begin{equation}
\psi_A(r) = \left(\frac{3}{2\pi}\right)^{3/4} \frac{1}{S_A^{3/2}}\, 
e^{-\left(\frac{3r^2}{4S_A^2}\right)}\;\,,
\end{equation}

where $S_A$ is the spread of the corresponding MLWF, then one can easily
obtain that :

\begin{equation}
O_{AB} = 8 \frac{S_A^3 S_B^3}{(S_A^2+S_B^2)^3}\,
e^{-\left(\frac{3}{2} \frac{R_{AB}^2}{S_A^2+S_B^2}\right)}\;\,.
\end{equation}

Then, the exchange integral can be expressed in terms of the
MLWFs spreads as :

\begin{equation}
J_{ex}^{AB}(R_{AB})= \frac{q^2 O_{AB}}{2R_{AB}} =
4 \frac{q^2}{R_{AB}} \frac{S_A^3 S_B^3}{(S_A^2+S_B^2)^3}\,
e^{-\left(\frac{3}{2} \frac{R_{AB}^2}{S_A^2+S_B^2}\right)}\;\,.
\end{equation}

Therefore, in this new DFT/vdW-WF2-x version of the method,
the vdW interaction energy is computed as:

\begin{equation}
E_{vdW}=-\sum_{i<j}\frac{C_6^{ij}}{R^6_{ij}} +
\sum_{i<j} J_{ex}^{ij}(R_{ij}) \;\,.
\label{esumnew}
\end{equation}

In this way the vdW energy correction is evaluated as the sum of
two terms, both expressed in terms of the MLWfs spreads, thus
making explicit the direct connection between attractive and repulsive
parts of the vdW interaction.\cite{Fedorov} 

Of course there are very weakly bonded systems, entirely dominated by
vdW effects, where the repulsive term is not relevant for determining
the equilibrium complex configuration. 
For instance, in the Ar-dimer case, 
DFT/vdW-WF2 and DFT/vdW-WF2-x predict the same equilibrium
Ar-Ar distance and the same binding energy (within 0.1 meV).
However, in most cases, a proper treatment of short-range repulsion is
crucial to correctly describe the minimum, equilibrium configuration.

The calculations have been performed 
with both the CPMD\cite{CPMD} and the Quantum-ESPRESSO ab initio 
package\cite{ESPRESSO}
(in the latter case the MLWFs have been generated as a post-processing 
calculation using the WanT package\cite{WanT}), using norm-conserving
or ultrasoft pseudopotentials to describe the electron-ion interactions 
and taking 
mainly PBE\cite{PBE} as the reference, Generalized Gradient Approximation 
(GGA) DFT functional, although test calculations have been also carried out
using the BLYP\cite{BLYP} GGA functional.
PBE and BLYP are chosen because they represent two of the most
popular GGA functionals for standard DFT simulations of condensed-matter
systems.

\section{Results and Discussion}
In order to assess the accuracy of the DFT/vdW-WF2-x method
we first consider the
S22 database of intermolecular interactions,\cite{Jurecka} a widely
used benchmark database, consisting of weakly interacting molecules
(a set of 22 weakly interacting dimers mostly of biological importance), 
with reference binding energies calculated
by a number of different groups using high-level
quantum chemical methods. In particular, we
use the basis-set extrapolated CCSD(T) binding
energies calculated by Takatani {\em et al}.\cite{Takatani} These binding
energies are presumed to have an accuracy of about 0.1
kcal/mol (1\% relative error). 
Calculations have been performed using
the same technical parameters adopted in ref. \onlinecite{myQHO}.

Table I summarizes
the results of our calculations on the S22 database, at the 
DFT/vdW-WF2-x level, considering PBE (DFT/vdW-WF2-x(PBE)) or BLYP
(DFT/vdW-WF2-x(BLYP)) as the reference DFT functional, compared to those
obtained by other vdW-corrected DFT schemes, namely 
DFT/vdW-WF2,\cite{C3}
vdW-DF,\cite{Dion,Langreth07} vdW-DF2,\cite{Lee-bis} VV10\cite{Vydrov} and
rVV10\cite{Sabatini}
(the revised, computationally much more efficient version
of the VV10 method), PBE+TS-vdW,\cite{TS} and PBE+MBD.\cite{Tkatchenko12}
For the sake of completeness we also report data relative to the semiempirical
PBE-D3\cite{Grimme} approach and to the bare, non-vdW-corrected, PBE 
and BLYP functionals.
In Table II, the performance of different schemes is illustrated
by separately considering {\it Hydrogen-bonded}, {\it dispersion}, and 
{\it mixed} complexes,
while Fig. 1 reports the behavior of the binding energy for all the
22 complexes contained in the S22 database, listed (for the sake of better
visibility) in the order of increasing (absolute) value of the 
reference, binding energy.
As expected, considering the whole S22 database, pure PBE and BLYP perform 
poorly and a substantial improvement can be obtained by vdW-corrected
approaches. More importantly, the performances of the new DFT/vdW-WF2-x 
scheme are clearly better than those of the previous DFT/vdW-WF2 method.
In particular, the general tendency of DFT/vdW-WF2 to overbind is considerably 
reduced by DFT/vdW-WF2-x.
Interestingly, with DFT/vdW-WF2-x(PBE) the mean absolute error (MAE),
0.78 kcal/mol, is well below the so-called "chemical accuracy" threshold
of 1 kcal/mol, required to attribute a genuine
quantitative character to the predictions of an ab initio scheme. 
Moreover, DFT/vdW-WF2-x(PBE) performs better than
the more sophisticated vdW-DF and vdW-DF2 methods, based on the use of
a nonlocal expression for the correlation energy-term, is comparable, as far 
as the S22 database is concerned to PBE-D3, and its
performances are only inferior to those of the rVV10, VV10, PBE+TS-vdW,
and PBE+MBD schemes, which are among the most accurate vdW-corrected 
DFT approaches for noncovalently bound complexes.\cite{Sabatini,Tkatchenko12} 
As can be seen, looking at Table II, DFT/vdW-WF2-x(PBE) turns out to be 
better than DFT/vdW-WF2-x(BLYP) for both dispersion-dominated and mixed
complexes, while instead the opposite is true for Hydrogen-bonded
complexes.

In order to test the applicability of the present DFT/vdW-WF2-x method also 
to a representative of extended systems, which of course is of particular
interest because, in this case, high-quality 
chemistry methods are typically too computationally demanding, 
we considered the adsorption of a single Ar atom on graphite, that is 
a typical physisorption process.
Calculations have been performed using
the same DFT approach followed in ref. \onlinecite{ambr} and considering
the adsorption on the more favored {\it hollow} site only.
Table III reports the binding energy, E$_b$, and equilibrium distance R,
for an Ar atom adsorbed on graphite, while in Fig. 2 the corresponding
binding energy curves are shown.
Data obtained at the PBE, DFT/vdW-WF2(PBE), and DFT/vdW-WF2-x(PBE) 
level are compared to reference theoretical and experimental estimates.
Theoretical values were obtained by
Tkatchenko {\it et al.},\cite{Tkatchenko} with a DFT-vdW corrected scheme
based on a semiempirical dispersion calibrated atom-centered potential,
and by Bichoutskaia and Pyper, with the inclusion of Axilord-Teller 
dispersion interactions.\cite{Ar}
As can be seen, the pure PBE functional largely underbinds while
vdW-corrected schemes predict a much stronger Ar-graphite interaction
with the formation of a clear minimum in the
binding energy curve at a shorter equilibrium distance.
At relatively large Ar-graphite distances, as expected, the 
DFT/vdW-WF2-x(PBE) curve approaches the DFT/vdW-WF2(PBE) one,
since the repulsive term becomes irrelevant. However, near the
equilibrium position, which is just determined by an interplay 
between attractive vdW interaction and repulsion, the differences
in the binding energies are significant, with DFT/vdW-WF2-x(PBE) which
represents an evident improvement compared to DFT/vdW-WF2(PBE), and,
in line with what previously observed in the application to the S22 database,
leads to a reduction of the binding-energy estimate, thus showing that the
effect of the repulsive term is more relevant. 
This can be seen more explicitly looking at Fig. 3, where the repulsive
contribution (see eq. \eqref{esumold}) of the DFT/vdW-WF2(PBE) method,
$\sum_{i<j}\left(1-f(R_{ij})\right)\frac{C_6^{ij}}{R^6_{ij}}$,
is compared to that (see eq. \eqref{esumnew}) of DFT/vdW-WF2-x(PBE),
$\sum_{i<j} J_{ex}^{ij}(R_{ij})$. Clearly, the repulsive term vanishes
for Ar-graphite distances larger than 4 \AA, however, around the 
equilibrium distance (and at shorter distances), it is much more
substantial in the DFT/vdW-WF2-x(PBE) approach than in DFT/vdW-WF2(PBE).

The DFT/vdW-WF2-x(PBE) binding energy (-115 meV) is very close 
to the values reported by Tkatchenko {\it et al.}\cite{Tkatchenko}
(-116 meV) and Bichoutskaia and Pyper\cite{Ar} (-111 meV). 
Moreover, this value is also compatible with
one of the few experimental estimates, represented
by the measurement of the latent heat of condensation relative to
the adsorption of an Ar monolayer on graphite: -119 $\pm$ 2 
meV/atom.\cite{Shaw} It is also 
close to the ''best estimate'' (obtained from a combination of
experimental and theoretical, mainly semiempirically-based,
data) reported in the milestone review paper by 
Vidali {\it et al.},\cite{Vidali} that is -99 $\pm$ 4 meV.
Our DFT/vdW-WF2-x(PBE) computed Ar-graphite equilibrium distance
is instead somehow larger than that reported in other theoretical 
studies\cite{Ar,Tkatchenko} (3.3 \AA), and also
than the ''best estimate'' by Vidali {\it et al.}\cite{Vidali} of 
3.1 $\pm$ 0.1 \AA\ (and an old experimental measurement of
3.2 $\pm$ 0.1 \AA, see ref. \onlinecite{Vidali}). However one sould observe
that an accurate estimate of this quantity is more difficult due
to the relatively shallow potential energy curve of this system.

\section{Conclusions}
In summary, we have presented a 
new method for including van der Waals (vdW) interactions
in Density Functional Theory using the MLWFs,
which is free from empirical parameters. 
With respect to the previous DFT/vdW-WF2 method, in the
present DFT/vdW-WF2-x approach, the empirical, short-range, damping 
function is replaced by an estimate of the Pauli exchange repulsion,
also obtained by the MLWFs properties.
Applications to systems contained in 
the popular S22 molecular database and to the case of an Ar atom interacting 
with graphite, and comparison with reference data,
indicate that the new method, besides being more physically founded, 
also leads to a systematic improvement in the description 
of vdW-bonded systems.

\section{Acknowledgements}
We acknowledge funding from Fondazione Cariparo, Progetti di Eccellenza 2017,
relative to the project: ''Engineering van der Waals
Interactions: Innovative paradigm for the control of Nanoscale
Phenomena''.

\vfill
\eject

\begin{table}
\caption{
Performance of different schemes
on the S22 database of intermolecular interactions.
The errors are measured with respect to basis-set
extrapolated CCSD(T) calculations of Takatani {\em et al}.\cite{Takatani}
Mean absolute relative errors (MARE in \%)
and mean absolute errors (MAE in
kcal/mol, and, in parenthesis, in meV) are reported.} 
\begin{center}
\begin{tabular}{|l|r|r|}
\hline
method & MARE & MAE \\ \tableline
\hline
PBE                 & 55.5 & 2.56[111.0] \\
DFT/vdW-WF2(PBE)    & 24.4 & 1.50 [65.1] \\
DFT/vdW-WF2-x(PBE)  & 13.4 & 0.78 [33.8] \\
\hline
BLYP                & 52.9 & 2.24 [97.1] \\
DFT/vdW-WF2(BLYP)   & 31.1 & 1.94 [84.1] \\
DFT/vdW-WF2-x(BLYP) & 20.5 & 1.24 [53.8] \\
\hline
vdW-DF$^a$          & 17.0 & 1.22 [52.9] \\
vdW-DF2$^b$         & 14.7 & 0.94 [40.8] \\
VV10$^b$            &  4.4 & 0.31 [13.4] \\ 
rVV10$^c$           &  4.3 & 0.30 [13.0] \\ 
PBE+TS-vdW$^{d,e}$  & 10.3 & 0.32 [13.9] \\
PBE+MBD$^d$         &  6.2 & 0.26 [11.3] \\
PBE-D3$^{c,f}$      & 11.4 & 0.50 [21.7] \\ 
\hline
\end{tabular}
\tablenotetext[1]{ref.\onlinecite{Cooper}.} 
\tablenotetext[2]{ref.\onlinecite{Vydrov}.} 
\tablenotetext[3]{ref.\onlinecite{Sabatini}.} 
\tablenotetext[4]{ref.\onlinecite{PNAS}.} 
\tablenotetext[5]{ref.\onlinecite{TS}.} 
\tablenotetext[6]{ref.\onlinecite{Grimme}.} 
\end{center}
\label{table1}
\end{table}
\vfill
\eject

\begin{table}
\caption{
Performance of different schemes
on the S22 database of intermolecular interactions, by separately considering
{\it Hydrogen-bonded}, {\it dispersion}, and {\it mixed} complexes.
The errors are measured with respect to basis-set
extrapolated CCSD(T) calculations of Takatani {\em et al}.\cite{Takatani}
Mean absolute relative errors (MARE in \%)
and mean absolute errors (MAE in
kcal/mol, and, in parenthesis, in meV) are reported.} 
\begin{center}
\begin{tabular}{|l|r|r|}
\hline
method & MARE & MAE \\ \tableline
\hline
{\it Hydrogen-bonded complexes}: &      &             \\
PBE                 &  8.4 & 1.22 [52.9] \\
DFT/vdW-WF2(PBE)    & 18.5 & 2.42[105.0] \\
DFT/vdW-WF2-x(PBE)  & 10.8 & 1.21 [52.5] \\
BLYP                & 12.6 & 1.53 [66.4] \\
DFT/vdW-WF2(BLYP)   & 14.3 & 2.11 [91.5] \\
DFT/vdW-WF2-x(BLYP) &  6.6 & 0.90 [39.0] \\
\hline
{\it Dispersion complexes}: &      &             \\
PBE                 &106.4 & 4.54[196.9] \\
DFT/vdW-WF2(PBE)    & 38.4 & 1.55 [67.2] \\
DFT/vdW-WF2-x(PBE)  & 21.0 & 0.85 [36.9] \\
BLYP                & 91.6 & 3.27[141.8] \\
DFT/vdW-WF2(BLYP)   & 58.1 & 2.84[123.2] \\
DFT/vdW-WF2-x(BLYP) & 41.0 & 2.15 [93.2] \\
\hline
{\it Mixed complexes}: &      &             \\
PBE                 & 51.6 & 2.00 [86.7] \\
DFT/vdW-WF2(PBE)    & 14.3 & 0.52 [22.6] \\
DFT/vdW-WF2-x(PBE)  &  7.2 & 0.25 [10.8] \\
BLYP                & 48.9 & 1.78 [77.2] \\
DFT/vdW-WF2(BLYP)   & 17.0 & 0.75 [32.5] \\
DFT/vdW-WF2-x(BLYP) & 11.0 & 0.52 [22.6] \\
\hline
\end{tabular}
\end{center}
\label{table2}
\end{table}
\vfill
\eject

\begin{table}
\caption{Binding energy, E$_b$, and equilibrium distance R,
for an Ar atom adsorbed on graphite.}
\begin{center}
\begin{tabular}{|l|r|c|}
\hline
method & E$_b$ (meV) &  R (\AA) \\ \tableline
\hline
PBE                 & -12       & 4.0 \\
DFT/vdW-WF2(PBE)    &-145       & 3.3 \\
DFT/vdW-WF2-x(PBE)  &-115       & 3.5 \\ 
\hline
ref.theory$^a$      &-116         & 3.3 \\
ref.theory$^a$      &-111         & 3.3 \\
ref.expt.$^c$       & -99 $\pm$ 4 & 3.0 $\pm$ 0.1 \\
ref.expt.$^d$       & ---         & 3.2 $\pm$ 0.1 \\
\hline
\end{tabular}
\tablenotetext[1]{ref.\onlinecite{Tkatchenko}.}
\tablenotetext[2]{ref.\onlinecite{Ar}.}
\tablenotetext[3]{ref.\onlinecite{Vidali}.}
\tablenotetext[4]{ref.\onlinecite{Vanselow}.}
\end{center}
\label{table3}
\end{table}
\vfill
\eject

\begin{figure}
\centerline{
\includegraphics[width=17cm,angle=270]{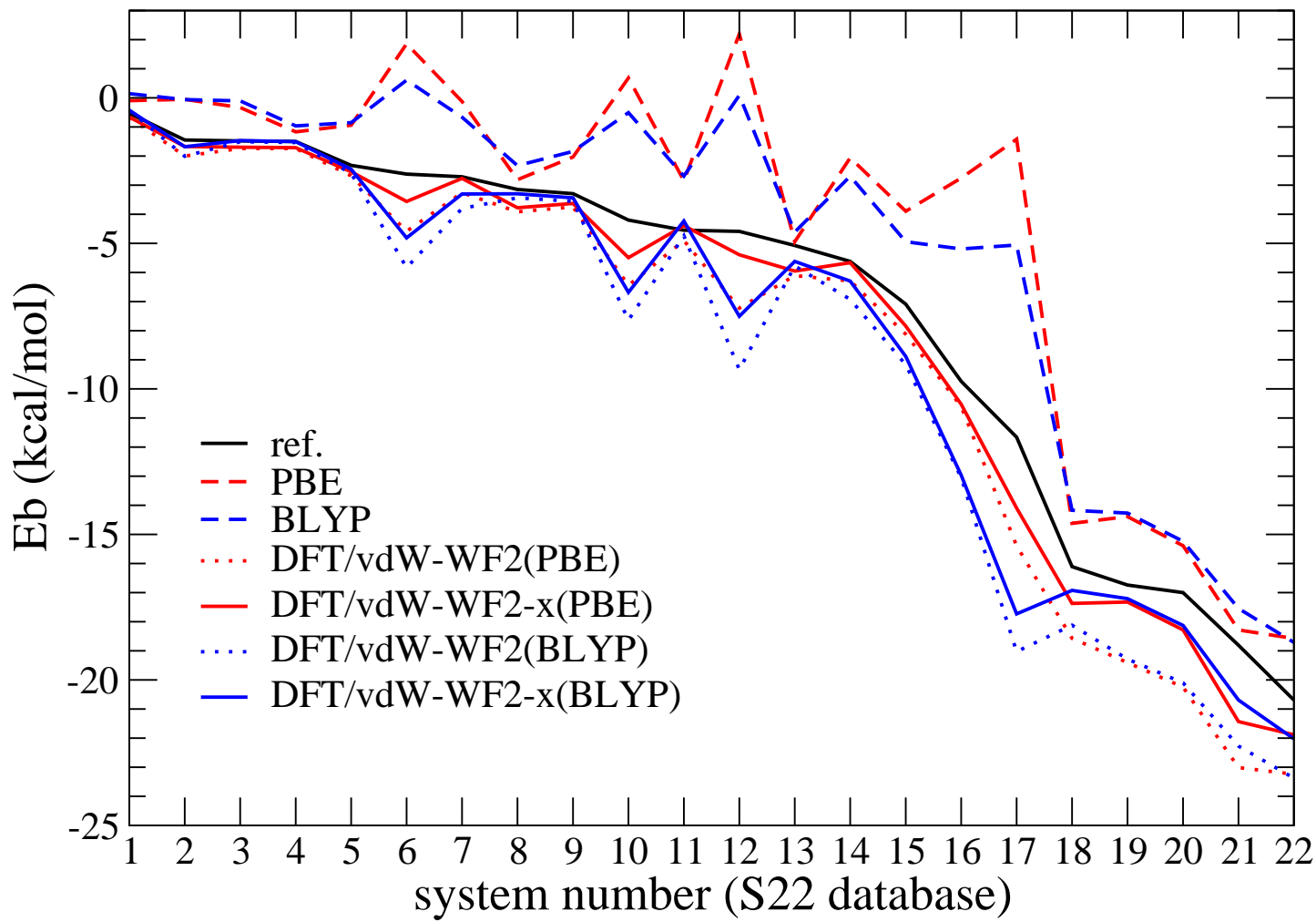}
}
\caption{Binding energy for all the
22 complexes contained in the S22 database, listed (for the sake of better
visibility) in the order of increasing (absolute) value of the 
reference binding energy using different functionals.}
\label{s22}
\huge
\end{figure}
\eject

\begin{figure}
\centerline{
\includegraphics[width=17cm,angle=270]{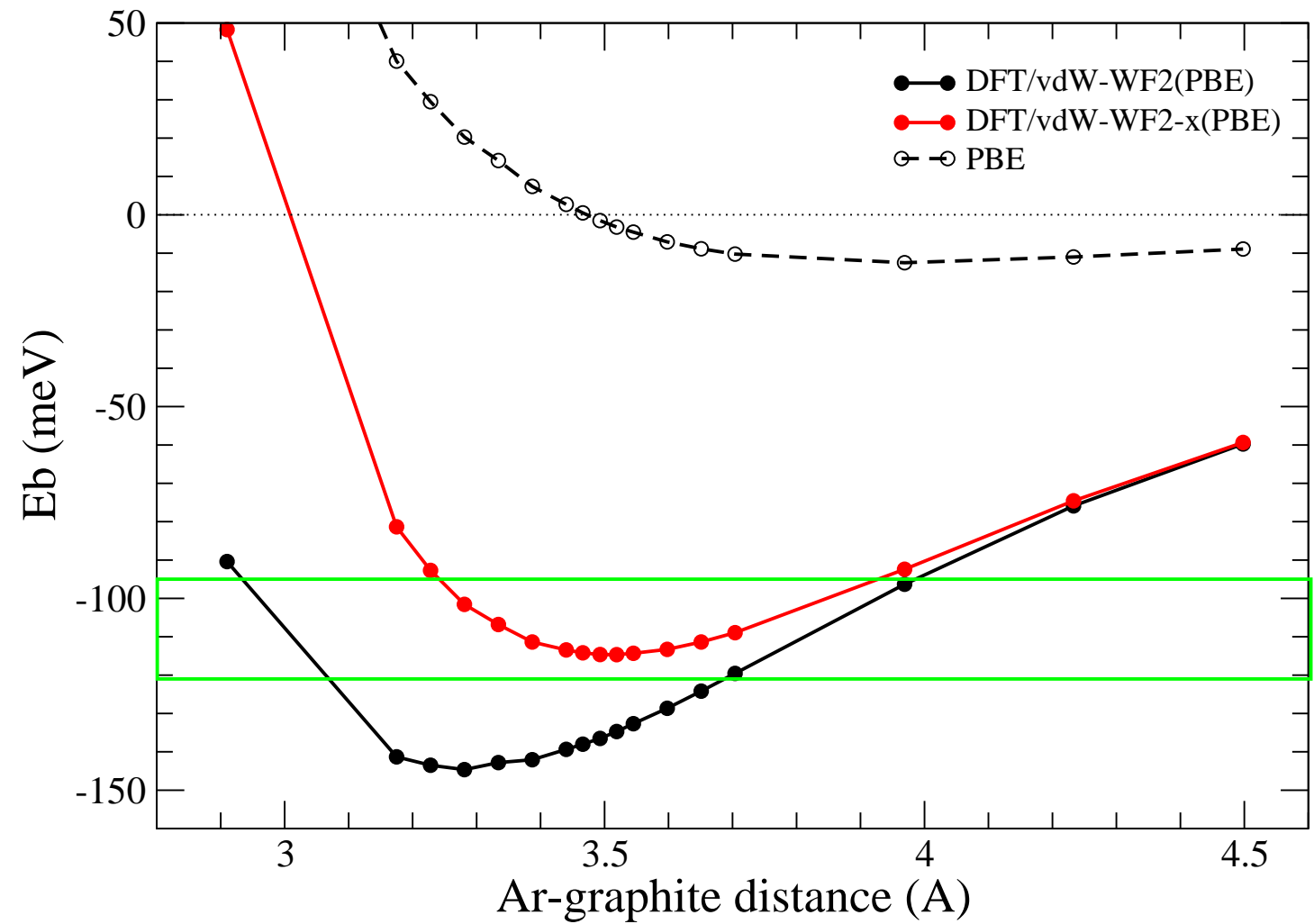}
}
\caption{Binding energy of Ar on graphite.
The horizontal green box denotes the binding-energy range corresponding
to the theoretical and experimental estimates (see Table III).}
\label{ar-gra}
\huge
\end{figure}
\eject

\begin{figure}
\centerline{
\includegraphics[width=17cm,angle=270]{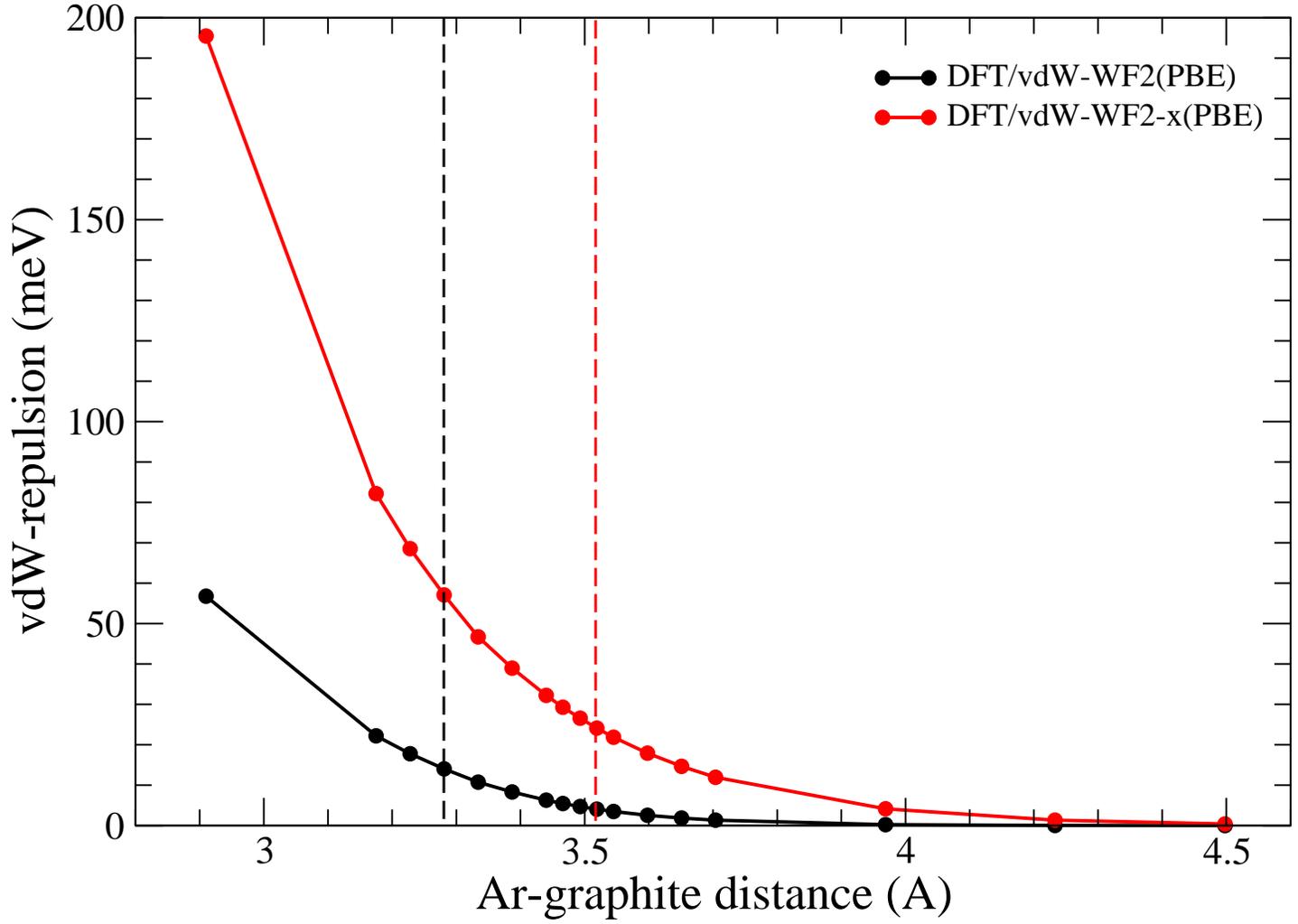}
}
\caption{Repulsive component of the vdW correction (see text) for 
Ar on graphite. The vertical dashed lines indicate the positions of
the equilibrium distances.}
\label{atrep}
\huge
\end{figure}
\eject

\end{document}